\documentclass[aps, prb, reprint, groupedaddress, superscriptaddress, showpacs]{revtex4-1}

\pdfoutput=1

\usepackage{graphicx}
\graphicspath{{./figs/}}
\usepackage{commath, amsmath, amssymb, amsfonts}
\usepackage{hyperref}
\usepackage{physics, siunitx}
\usepackage{epstopdf}

\usepackage[normalem]{ulem}
\usepackage[usenames,dvipsnames]{xcolor}

\renewcommand{\va}[1]{\mathbf{#1}}
\newcommand{\scase}[1]{\mathcal{{#1}}}
\newcommand{\field}[1]{\boldsymbol{\mathcal{{#1}}}}
\usepackage{array}
\newcolumntype{P}[1]{>{\centering\arraybackslash}p{#1}}
\newcolumntype{M}[1]{>{\centering\arraybackslash}m{#1}}
\allowdisplaybreaks
\newcommand{\ie}{i.e.~}
\newcommand{\cf}{c.f.~}
\newcommand{\eg}{e.g.~}

\usepackage{mathtools}
\DeclarePairedDelimiter{\evdel}{\langle}{\rangle}
\renewcommand{\ev}{\evdel}

\hypersetup{
	colorlinks   = true, 
	urlcolor     = blue, 
	linkcolor    = blue, 
	citecolor   = blue 
}

\begin{document}


\title{Gauge invariance of excitonic linear and nonlinear optical response}


\author{Alireza Taghizadeh}
\email{ata@nano.aau.dk}
\affiliation{Department of Physics and Nanotechnology, Aalborg University, DK-9220 Aalborg {\O}st, Denmark}

\author{T. G. Pedersen}
\affiliation{Department of Physics and Nanotechnology, Aalborg University, DK-9220 Aalborg {\O}st, Denmark}
\affiliation{Center for Nanostructured Graphene (CNG), DK-9220 Aalborg {\O}st, Denmark}
\date{(Received XX XXXXXX 2018, published XX XXXXXX 2018}%


\date{\today}

\begin{abstract}
We study the equivalence of four different approaches to calculate the excitonic linear and nonlinear optical response of multiband semiconductors. These four methods derive from two choices of gauge, \ie length and velocity gauges, and two ways of computing the current density, \ie direct evaluation and evaluation via the time-derivative of the polarization density. The linear and quadratic response functions are obtained for all methods by employing a perturbative density matrix approach within the mean-field approximation. The equivalence of all four methods is shown rigorously, when a correct interaction Hamiltonian is employed for the velocity gauge approaches. The correct interaction is written as a series of commutators containing the unperturbed Hamiltonian and position operators, which becomes equivalent to the conventional velocity gauge interaction in the limit of infinite Coulomb screening and infinitely many bands. As a case study, the theory is applied to hexagonal boron nitride monolayers, and the linear and nonlinear optical response found in different approaches are compared.
\end{abstract}


\maketitle

\section{Introduction \label{sec:Introduction}}

The optical response of crystals provides valuable information about material properties, \eg important features of the band structure \cite{Boyd2008, Yu2010}. The response can be characterized by the linear response as well as diverse nonlinear ones, \eg second/third harmonic generation, optical rectification, etc\cite{Boyd2008}. Theoretically, accurate estimates of optical response functions based on the material band structure are highly desirable, since they can offer important insights for experiments and device applications. 
Nowadays, perturbative calculations of linear and nonlinear optical response functions are routinely performed in the independent-particle approximation (IPA), in which the electron-hole interaction is simply ignored, \eg see Refs.~\onlinecite{Sipe1993, Aversa1995, Rashkeev2001, Pedersen2009, Margulis2013, Mikhailov2016, Hipolito2016, Hipolito2018, Julen2018} (and references therein).
However, it is well-known that including the electron-hole interaction, \ie excitonic effects, can have a significant influence on the optical response of solids \cite{Leung1997, Albrecht1998, Benedict1998, Rohlfing2000, Onida2002}. In particular, excitons dramatically modify the optical response of low-dimensional systems including carbon nanotubes \cite{Pedersen2003b, Wang2005}, and two-dimensional (2D) materials such as hexagonal boron nitride (hBN) \cite{Pedersen2015, Cudazzo2016, Galvani2016, Koskelo2017} and transition metal dichalcogenides \cite{Gruning2014, Trolle2015, Gao2016, Qiu2016, Olsen2016} due to the reduced screening and enhanced confinement of electrons. Typically, excitons affect the linear response by introducing strong resonances inside the band gap and renormalizing the continuum part of spectrum\cite{Pedersen2015, Attaccalite2017}. Regarding the nonlinear optical response, a few theoretical studies have been done both on bulk \cite{Chang2001, Leitsmann2005} and 2D materials \cite{Gruning2014, Trolle2014, Pedersen2015}, which show that complex modifications of the spectra occur due to excitons.  


From a theoretical point of view, the optical response should, in principle, be independent of the chosen electromagnetic gauge. However, in practice, the choice of electromagnetic gauge, \eg the so-called length and velocity gauges, influences the results due to various approximations \cite{Kobe1979, Girlanda1981, Madsen2002, Rzazewski2004, Virk2007, Dong2014, Foldi2017, Ventura2017, Taghizadeh2017, Yakovlev2017}. 
Formally, it is straightforward to show that the wavefunctions in the length and velocity gauges are related via a time-dependent unitary transformation \cite{Kobe1979}. The unitary transformation converts the length gauge (LG) Hamiltonian to its equivalent velocity gauge (VG) counterpart. This Hamiltonian includes the unperturbed Hamiltonian plus an interaction part, which is given by a series of commutators between the position and unperturbed Hamiltonian \cite{Girlanda1981, Passos2017}. In the IPA limit, the series can be truncated to the first two terms if the canonical commutator relation between position $\hat{\va{r}}$ and momentum $\hat{\va{p}}$ is used, \ie $[\hat{\va{r}},\hat{\va{p}}]=i\hbar\va{I}$ where $\va{I}$ denotes the unit tensor \cite{Girlanda1981}. This leads to the conventional VG (CVG) interaction given by $e\field{A}\vdot\hat{\va{p}}/m +e^2\field{A}^2/2m$, where $\field{A}$ is the vector potential. For an infinite periodic system, the calculation using this interaction has the advantage over the LG that it avoids using the ill-defined matrix elements of position and uses only the well-defined momentum ones. Nonetheless, the price paid for calculating the optical response in the CVG is that a large number of bands is typically required to obtain an acceptable result \cite{Taghizadeh2017}. Thus, if an insufficient number of bands is used in the calculations, the response functions computed in the CVG may suffer from the well-known zero-frequency divergences \cite{Sipe1993}, or even become entirely incorrect. For instance, the even-order responses obtained within a two-band model are identically zero due to the time-reversal symmetry \cite{Hipolito2016, Taghizadeh2017}. In contrast, if the exact series of commutators is used for the VG interaction, the LG and VG results become identical regardless of the number of bands in the calculations \cite{Passos2017}. However, one of the most important strength of the VG, which is its simple implementation, is utterly lost. 
 


Including the electron-hole interaction complicates the optical response calculation dramatically, which compels us to introduce additional approximations such as the mean-field approximation (MFA) to solve the many-body problem. Within the MFA, the Hamiltonian becomes effectively nonlocal due to the Coulomb term. Despite the nonlocality of the Hamiltonian, the MFA theory has been shown to be formally invariant under the gauge transformation \cite{Kobe1979}. Nonetheless, the meaning of the gauge freedom in the excitonic optical response calculation is not fully understood. For instance, using the CVG interaction, as in Refs.~\onlinecite{Trolle2014, Ridolfi2018}, results in an incorrect excitonic optical response regardless of the basis completeness as will be demonstrated in Sec.~\ref{sec:Gauge}. In addition to the gauge freedom, it is well-known that the optical response can be computed in two ways: direct evaluation of the current density and via the time-derivative of the polarization density \cite{Pedersen2015, Taghizadeh2017}. Similarly to the equivalence of the two gauges, these two approaches should, in principle, be equivalent but may generate different results when excitons are considered \cite{Pedersen2015}. Therefore, four formally-equivalent but computationally-different methods for obtaining the optical response are available, which consist of the combinations of two choices of gauge, \ie LG and VG, and two ways of computing the current density: directly and indirectly via the polarization density. 


In the present work, we investigate systematically the influence of the chosen gauge and observable on the excitonic linear and nonlinear optical response. We develop a practical framework for calculating the multiband semiconductor response by adopting a many-body density matrix approach within the MFA. Employing a perturbative solution for the density matrix, the expressions of the first and second-order response functions are derived. Using this framework, we show that 1) the excitonic responses obtained using the four above-mentioned alternatives are identical regardless of the number of bands in the calculations if the VG interaction is written as the commutator series of the position and unperturbed Hamiltonian; 2) despite the equivalence of the four methods, it is simpler to derive the conductivity expressions and perform the calculations in the LG approaches compared to their VG counterparts, particularly in the case of nonlinear responses; 3) the excitonic response computed using the CVG interaction is not reliable in the MFA even if a complete basis is employed for the calculations, since the MFA Hamiltonian includes an effective nonlocal potential; 4) the response generated by the CVG and VG become identical if a complete basis set is used and the Coulomb interaction is neglected. We apply the proposed theory to hBN monolayers as a case study, and confirm the validity of the theoretical framework through numerical simulations. 





\section{Theoretical framework \label{sec:Theory}}

In this section, we present the theoretical framework for calculating the excitonic optical response of periodic systems. We begin by introducing the equation of motion for the density matrix. Then, using a perturbative solution of the dynamical equations, the linear and second-order optical conductivities are derived for all methods. Finally, the equivalence of response functions obtained by these different methods is discussed. It should noted that throughout the text, all vectors and tensors are indicated by bold letters, and the single-particle/many-body operators and matrix elements are denoted by lowercase/uppercase letters, respectively.

\subsection{Dynamical equation \label{sec:Dynamical}} 
The many-body Hamiltonian of a system of electrons under the influence of an external perturbation is written in second quantization as 
\begin{align}
	\label{eq:TotalHamiltonian}
	\hat{H} &\equiv \hat{H}_0 + \mathcal{\hat{V}} + \hat{U}(t) \equiv \mathcal{\hat{H}}_0 + \hat{U}(t) \equiv \sum_{k} \varepsilon_k^0 \hat{c}_k^\dagger \hat{c}_k \nonumber \\
	&+ \dfrac{1}{2} \sum_{klmn} \mathcal{V}_{klmn} \hat{c}_k^\dagger \hat{c}_l^\dagger \hat{c}_n \hat{c}_m + \sum_{kl} u_{kl}(t) \hat{c}_k^\dagger \hat{c}_l \, ,
\end{align}
where $\hat{c}$ and $\hat{c}^\dagger$ are the fermionic annihilation and creation operators, respectively, and $\hat{H}_0$, $\mathcal{\hat{V}}$, and $\hat{U}(t)$ are the single-electron, Coulomb potential, and time-dependent interaction parts, respectively. $\mathcal{\hat{H}}_0 \equiv \hat{H}_0 + \mathcal{\hat{V}}$ is the total unperturbed Hamiltonian. For simplicity, the spin-orbit coupling is neglected here. Note that the Hamiltonian $\hat{H}$ here is written in the single-particle basis $\ket{n}$, \ie $\hat{h}_0\ket{n}=\varepsilon_n^0\ket{n}$ with $\hat{h}_0$ the unperturbed Hamiltonian of a single electron, and the matrix elements of the external potential read $u_{kl}(t) \equiv \mel{k}{\hat{u}(t)}{l}$, where $\hat{u}(t)$ is the interaction potential of an individual electron.

The dynamical behavior of the system is then studied by employing a density matrix approach, for which we follow the procedure outlined in Ref.~\onlinecite{Pedersen2015} and explained in Appendix~\ref{sec:AppendixA}. Within the MFA, the equation of motion for the density matrix, $\rho_{ji}(t) \equiv \ev{\psi_0|\hat{c}_i^\dagger \hat{c}_j|\psi_0}$, is derived as shown in Eq.~(\ref{eq:DensityGeneral}). Here, $|\psi_0\rangle$ is the many-body ground state, in which all the valence states are occupied.
For the special case of periodic systems, the single-particle basis states are of the Bloch form $|n\va{k}\rangle=A^{-1/2}e^{i\va{k} \vdot \va{r}} \varphi_{n\va{k}}(\va{r})$, where $A$, $\varphi_{n\va{k}}(\va{r})$, $n$ and $\va{k}$ are the crystal volume, cell-periodic part, band index and wavevector, respectively. For a general single-particle operator $\hat{o}$, we denote single-particle matrix elements by $o_{nm\va{k}}\equiv\ev{n\va{k}|\hat{o}|m\va{k}}$ such as $\va{p}_{nm\va{k}}$ for the momentum.
In the Bloch basis, the equation of motion for the density matrix $\rho_{ji\va{k}}$ is given in Eq.~(\ref{eq:DensityInit}), which can be solved perturbatively up to any required order of perturbation.
The solutions for the first and second order are presented in Eqs.~(\ref{eq:DensityFinalA})-(\ref{eq:DensityFinalE}). These expressions are obtained in the so-called Tamm-Dancoff approximation \cite{Benedict1998, Rohlfing2000}, where the coupling between off-diagonal elements $\rho_{cv\va{k}}$ and $\rho_{vc\va{k}}$ is ignored (the indices $c$ and $v$ imply conduction and valence bands, respectively). Upon determining the density matrix, the expectation value of any observable of the system is found using Eqs.~(\ref{eq:ObservableFinalA}) and (\ref{eq:ObservableFinalB}) in Appendix~\ref{sec:AppendixA}. 


\subsection{Linear and quadratic optical response \label{sec:Optical}}
The optical response is calculated as the induced current density inside the material owing to the interaction with an external electromagnetic field. 
Throughout this work, the electric field $\field{E}$ is decomposed into its harmonic components, 
\begin{equation}
	\field{E}(t) = \dfrac{1}{2} \sum_p \field{E}(\omega_p) e^{-i\omega_p t} \, ,
\end{equation} 
where the $p$-summation is performed over both positive and negative frequencies. Note that we neglect the spatial variation of the field, \ie a long-wavelength regime is assumed.
On the other hand, the form of time-dependent interaction $\hat{U}(t)$ depends on the choice of gauge. In the LG, $\hat{U}_\textrm{LG}(t)=e\hat{\va{R}} \vdot \field{E}(t)$, where $\hat{\va{R}}$ denotes the many-body position operator. In the CVG, the interaction reads $\hat{U}_\textrm{CVG}(t)=e(\field{A} \vdot \hat{\va{P}} +egN\field{A}^2/2)/m$, where $gN$ and $\hat{\va{P}}$ are the total number of electrons and many-body momentum operator, respectively, and $g=2$ accounts for the spin degeneracy. Moreover, the vector potential  $\field{A}$ is mapped to $\field{E}$ via $\field{E}=-\partial \field{A}/\partial t$. When the electron-hole interaction is considered in the MFA, an effective nonlocal potential is introduced in the unperturbed Hamiltonian, which puts the validity of $\hat{U}_\textrm{CVG}(t)$ into question \cite{Kobe1979, Girlanda1981}. In this case, it has been shown in Refs.~\onlinecite{Starace1971, Girlanda1981} that the (correct) VG interaction should instead be written as a series of commutators. Up to the second-order in $\field{A}$, the interaction reads $\hat{U}_\textrm{VG}(t)=e(\field{A}\vdot\hat{\boldsymbol{\Pi}} +e [\field{A}\vdot\hat{\va{R}}, \field{A}\vdot\hat{\boldsymbol{\Pi}}]/2i\hbar)/m$, where $\hbar\hat{\boldsymbol{\Pi}} \equiv im[\mathcal{\hat{H}}_0, \hat{\va{R}}] \neq \hbar\hat{\va{P}}$. We refer to $\hat{\boldsymbol{\Pi}}$ as the Heisenberg momentum operator, since it is proportional to the time-derivative of the position operator (or velocity \cite{Girlanda1981}) in the Heisenberg picture, \ie $\dd \hat{\va{R}}/\dd t=i[\mathcal{\hat{H}}_0,\hat{\va{R}}]/\hbar$.  


In addition to the gauge freedom, it is possible to calculate the optical response either by evaluating directly the expectation value of the current density operator, $\va{J}(t) \equiv \ev{\hat{\va{J}}}$, or by computing the time-derivative of the expectation value of the polarization density operator, $\va{J}(t) \equiv \partial \boldsymbol{\mathcal{P}}(t)/\partial t = \partial \ev{\hat{\boldsymbol{\mathcal{P}}}}/\partial t$ \cite{Pedersen2015, Taghizadeh2017}. The many-body current and polarization density operators read $\hat{\va{J}} = -e\hat{\va{\Pi}}/(mA)$ and $\hat{\boldsymbol{\mathcal{P}}} = -e\hat{\va{R}}/A$, respectively, and the expectation values of the operators are determined by employing the density matrix as discussed in Sec.~\ref{sec:Dynamical}. Note that the current density operator in the VG includes an extra diamagnetic term and reads $\hat{\va{J}} = -e(\hat{\va{\Pi}} - e [\field{A}\vdot\hat{\va{R}},\hat{\va{\Pi}}]/i\hbar)/(mA)$. 
Hence, a total of four alternatives for computing the optical response are possible, which are formed by the combination of two gauges and two ways of evaluating the current density response as labeled in Table~\ref{tb:Gauges}. For comparison purposes, we include the CVG labeled by $\scase{C}'$, where $\hat{U}_\textrm{CVG}(t)$ is used as the interaction Hamiltonian.
Note that hereinafter, the normalized position operators $\hat{\va{x}}$ and $\hat{\va{X}}$ are used for convenience, which are defined as $\hat{\va{x}}\equiv m\hat{\va{r}}/\hbar$, and similarly for the many-body operator $\hat{\va{X}}\equiv m\hat{\va{R}}/\hbar$. Using the normalized position, the many-body canonical commutator relation becomes $[\hat{\va{X}},\hat{\va{P}}]=igNm\va{I}$.

\begin{table}[t]
	\caption[Methods to compute the current response]{
		\label{tb:Gauges}
		Four equivalent methods ($\scase{A}$-$\scase{D}$) for computing the current density response and their respective labels. For comparison purposes, the conventional velocity gauge method is also shown and labeled by $\scase{C}'$. Here, $\hat{\va{R}}$, $\hat{\va{P}}$, $\field{E}$, and $\field{A}$ represent the many-body position operator, many-body momentum operator, electric field, and vector potential, respectively, $g=2$ for spin degeneracy, and $gN$ is the total number of electrons. $\hat{\va{\Pi}}$ denotes the Heisenberg momentum operator defined as $\hbar\hat{\va{\Pi}} \equiv im[\hat{\mathcal{H}}_0,\hat{\va{R}}]$, where $\hat{\mathcal{H}}_0$ is the unperturbed Hamiltonian. } 
	\centering
	\begin{tabular}{p{0.7cm}p{3.5cm}p{3.1cm}}
		\hline
		Label & $\hat{U}(t) \propto $ &  $\va{J}(t) \propto $ \\  
		\hline
		$\scase{A}$ & $ \hat{\va{R}}\vdot\field{E}$ & $\ev{\hat{\va{\Pi}}}$ \\
		$\scase{B}$ & $ \hat{\va{R}}\vdot\field{E}$ & $\partial \ev{\hat{\va{R}}}/\partial t$ \\
		$\scase{C}$ & $\field{A}\vdot\hat{\va{\Pi}} +e[\field{A}\vdot\hat{\va{R}}, \field{A}\vdot\hat{\va{\Pi}}]/2i\hbar$ & $\ev{\hat{\va{\Pi}}+e [\field{A}\vdot\hat{\va{R}},\hat{\va{\Pi}}]/i\hbar}$ \\ 
		$\scase{D}$ & $\field{A}\vdot\hat{\va{\Pi}} +e[\field{A}\vdot\hat{\va{R}}, \field{A}\vdot\hat{\va{\Pi}}]/2i\hbar$ & $\partial \ev{\hat{\va{R}}}/\partial t$   \\
		$\scase{C}'$ & $\field{A}\vdot\hat{\va{P}} +egN \field{A}^2/2$ & $\ev{\hat{\va{P}} +egN\field{A}} $ \\
		\hline
	\end{tabular}
\end{table}

Without loss of generality, the first-order current density $\va{J}^{(1)}(t)$ reads
\begin{equation}
	\va{J}^{(1)}(t) = \dfrac{1}{2} \sum_p \boldsymbol{\sigma}^{(1)}(\omega_p) \field{E}(\omega_p) e^{-i\omega_p t} \, ,
\end{equation}
where the optical conductivity (OC) tensors $\boldsymbol{\sigma}^{(1)}(\omega_p)$ for the five methods of Table~\ref{tb:Gauges} are given by 
\begin{subequations}
	\label{eq:TensorFirst}
	\begin{align}
		\label{eq:TensorFirstA}
		&\boldsymbol{\sigma}^{\scase{A}(1)} = -C_e \sum_n \bigg[ \dfrac{\va{\Pi}_n \va{X}_n^*}{\hbar\omega_p-E_n} - \dfrac{\va{\Pi}_n^* \va{X}_n}{\hbar\omega_p+E_n} \bigg] \, , \\
		\label{eq:TensorFirstB}
		&\boldsymbol{\sigma}^{\scase{B}(1)} = C_e (i\hbar\omega_p) \sum_n \bigg[ \dfrac{\va{X}_n \va{X}_n^*}{\hbar\omega_p-E_n} - \dfrac{\va{X}_n^* \va{X}_n}{\hbar\omega_p+E_n} \bigg] \, , \\
		\label{eq:TensorFirstC}
		&\boldsymbol{\sigma}^{\scase{C}(1)} = \dfrac{i C_e}{\hbar\omega_p} \bigg\{ \sum_n \bigg[ \dfrac{\va{\Pi}_n \va{\Pi}_n^*}{\hbar\omega_p-E_n} - \dfrac{\va{\Pi}_n^* \va{\Pi}_n}{\hbar\omega_p+E_n} \bigg] + mN\va{L} \bigg\} \, , \\
		\label{eq:TensorFirstD}
		&\boldsymbol{\sigma}^{\scase{D}(1)} = \big[ \boldsymbol{\sigma}^{\scase{A}(1)} \big]^T \, , \\
		\label{eq:TensorFirstE}
		&\boldsymbol{\sigma}^{\scase{C}'(1)} = \dfrac{i C_e}{\hbar\omega_p} \bigg\{ \sum_n \bigg[ \dfrac{\va{P}_n \va{P}_n^*}{\hbar\omega_p-E_n} - \dfrac{\va{P}_n^* \va{P}_n}{\hbar\omega_p+E_n} \bigg] + mN\va{I} \bigg\} \, .
	\end{align}
\end{subequations}
Here, $C_e \equiv ge^2\hbar/(m^2 A)$, $\va{L} \equiv 2i\sum_n \va{\Pi}_n \va{X}_n^*/mN $, and $T$ denotes transposition. $E_n$ is the exciton energy obtained by solving the Bethe-Salpeter equation (BSE), \ie $H_{eh} |\psi^{(n)}\rangle = E_n |\psi^{(n)}\rangle$ with $H_{eh}$ given in Eq.~(\ref{eq:CouplingMatrix}). In addition, $\va{\Pi}_n = -iE_n\va{X}_n$ due to the definition of Heisenberg momentum $\hat{\va{\Pi}}$, and the excitonic momentum $\va{P}_n$ and position $\va{X}_n$ are defined as 
\begin{align}
	&\va{P}_n \equiv \sum_{cv\va{k}} \psi_{cv\va{k}}^{(n)} \va{p}_{vc\va{k}} \, , \, \, \va{X}_n \equiv \sum_{cv\va{k}} \psi_{cv\va{k}}^{(n)} \va{x}_{vc\va{k}} \, , 
\end{align}
where $\psi_{cv\va{k}}^{(n)}=\langle v\va{k}{\rightarrow}c\va{k} | \psi^{(n)} \rangle $ is the exciton projection onto a singlet band-to-band transition, and the summation over $\va{k}$ implies an integral over the first Brillouin zone (BZ), \ie $(2\pi)^D \sum_{\va{k}} \rightarrow A \int_{\mathrm{BZ}} \dd[D]{\va{k}}$ ($D=2$ for 2D materials). Note that $\va{P}_n$, $\va{X}_n$, and $\va{\Pi}_n$ are indeed the matrix elements of the many-body momentum, position and Heisenberg momentum operators between the ground state $|\psi_0\rangle$ and excited state $|\psi^{(n)}\rangle$, respectively.

Similarly, the quadratic current density response reads
\begin{equation}
	\va{J}^{(2)}(t) = \dfrac{1}{4} \sum_{p,q} \boldsymbol{\sigma}^{(2)}(\omega_p,\omega_q) \field{E}(\omega_p) \field{E}(\omega_q) e^{-i(\omega_p+\omega_q) t} \, ,
\end{equation}
where $\boldsymbol{\sigma}^{(2)}(\omega_p,\omega_q)$ are rank-3 conductivity tensors given for the five methods in Eqs.~(\ref{eq:TensorSecondA})-(\ref{eq:TensorSecondE}). The conductivities are written in terms of matrix elements for transition between two excitons $n$ and $m$ denoted by $O_{nm}=\ev{\psi^{(n)}|\hat{O}|\psi^{(m)}}$ such as $\va{X}_{nm}$. We note that expressions for optical susceptibilities $\boldsymbol{\chi}$ can readily be derived from their corresponding conductivities by $\epsilon_0\boldsymbol{\chi}^{(1)}\equiv i\boldsymbol{\sigma}^{(1)}/\omega_p$ and $\epsilon_0\boldsymbol{\chi}^{(2)}\equiv i\boldsymbol{\sigma}^{(2)}/(\omega_p+\omega_q)$. The derivation of the quadratic conductivity tensor in the CVG, \ie Eq.~(\ref{eq:TensorSecondE}), is a rather straightforward problem, since it only contains the well-defined matrix elements of momentum. In contrast, the intraband part of the single-particle position operator appearing in $\va{X}_{nm}$, Eq.~(\ref{eq:PositionMatEl}), is ill-defined inherently for infinite periodic systems. 

In spite of the problems associated with the position operator, it has been shown in Ref.~\onlinecite{Aversa1995} that the optical response can be computed by separating formally the interband and intraband parts of the position matrix elements, \ie $\ev{n\va{k}|\hat{\va{x}}|m\va{k}}=\ev{n\va{k}|\hat{\va{x}}^{(i)}|m\va{k}}\delta_{nm}+\ev{n\va{k}|\hat{\va{x}}^{(e)}|m\va{k}}(1-\delta_{nm})$. The interband part is simply related to the momentum matrix element \cite{Hipolito2016}, whereas the intraband block is handled by employing a commutator relation \cite{Aversa1995}:
\begin{subequations}
	\label{eq:GenDerivative}
	\begin{align}
		&\dfrac{\hbar}{m}\mel{n\va{k}}{[\hat{\va{x}}^{(i)},\hat{o}]}{m\va{k}} = i (o_{nm\va{k}})_{;\va{k}} \, , \\
		&(o_{nm\va{k}})_{;\va{k}} \equiv \gradient_{\va{k}} o_{nm\va{k}} - i[\va{\Omega}_{nn\va{k}} - \va{\Omega}_{mm\va{k}}] o_{nm\va{k}} \, .
	\end{align}
\end{subequations}
Here, $(o_{nm\va{k}})_{;\va{k}}$ is the generalized derivative written in terms of the Berry connections $\va{\Omega}_{nm\va{k}} \equiv iA_\textrm{uc}^{-1}\int_\textrm{uc} \varphi_{n\va{k}}^*(\va{r}) \gradient_{\va{k}} \varphi_{m\va{k}}(\va{r}) \dd[D]{\va{r}}$ ($A_\textrm{uc}$ is the unit-cell volume). By employing this technique, the interband/intraband parts of the position operator in $\va{X}_{nm}$ are separated. This separation leads to $\va{X}_{nm} = \va{Y}_{nm} + m\va{Q}_{nm}/\hbar$, where $\va{Y}_{nm}$ [see Eq.~(\ref{eq:InterbandPosition})] and $\va{Q}_{nm}$ [see Eq.~(\ref{eq:IntrabandPosition})] are the interband and intraband parts of excitonic position matrix elements, respectively.
Therefore, the quadratic conductivities in Eqs.(\ref{eq:TensorSecondA})-(\ref{eq:TensorSecondD}) consist of two distinct blocks: an interband contribution (terms including $\va{Y}_{nm}$) and an intraband part (terms containing $\va{Q}_{nm}$). Despite the seemingly distinct appearance of Eqs.~(\ref{eq:TensorSecondA}) to (\ref{eq:TensorSecondD}), they are equivalent as illustrated analytically in Sec.~\ref{sec:Gauge} and numerically in Sec.~\ref{sec:Results}.
We note that the VG quadratic conductivities computed using $\hat{U}_\textrm{VG}$ require the evaluation of $\va{Q}_{nm}$, \ie the generalized derivative, which is in contrast to the CVG using $\hat{U}_\textrm{CVG}$. Hence, the main advantage of performing computation in the VG, which is the absence of the generalized derivative, is lost if $\hat{U}_\textrm{VG}$ is used. In fact, computing the nonlinear conductivities in the VG with the correct interaction Hamiltonian, \ie tensors labeled by $\scase{C}$ and $\scase{D}$, is more complicated than the LG approaches, due to the presence of several extra terms in the conductivity expressions [\cf Eqs.~(\ref{eq:TensorSecondC}) and (\ref{eq:TensorSecondD})]. This becomes even more difficult in higher-order nonlinear responses due to additional terms in the interaction Hamiltonian and observable. Finally, as will be emphasized in Sec.~\ref{sec:Results}, a dense $\va{k}$-vector grid is typically essential in order to eliminate the apparent zero-frequency divergence of $\boldsymbol{\sigma}^{\scase{C}(2)}$ and $\boldsymbol{\sigma}^{\scase{D}(2)}$.



\subsection{Gauge invariance \label{sec:Gauge}}
It is straightforward to show that the initial dynamical equation for the density matrix in the MFA, \ie Eq.~(\ref{eq:DensityGeneral}), behaves in a gauge-independent manner. Therefore, it is expected that the ultimate expressions for the linear and nonlinear optical response, \ie Eqs.~(\ref{eq:TensorFirst}) and (\ref{eq:TensorSecond}), are equivalent. Indeed, we demonstrate in this section that the expressions obtained by methods $\scase{A}$ to $\scase{D}$ are equivalent. Regarding the CVG approach using $\hat{U}_\textrm{CVG}$, \ie tensors labeled by $\scase{C}'$, we show that they are generally different from the rest.

First, let us focus on the tensors labeled by $\scase{A}$ to $\scase{D}$. Beginning with the linear response, it is obvious that the OC tensors obtained using methods $\scase{A}$, $\scase{B}$ and $\scase{D}$ are indeed identical, since $\va{\Pi}_n=-iE_n\va{X}_n$. For the method $\scase{C}$, the denominators in Eq.~(\ref{eq:TensorFirstC}) are decomposed using a partial fraction expansion and rewritten as $1/\hbar\omega(\hbar\omega \pm E_n)=\pm1/E_n(\hbar\omega \pm E_n)\mp1/\hbar\omega E_n$. The terms due to $1/E_n(\hbar\omega \pm E_n)$ form a conductivity identical to the $\boldsymbol{\sigma}^{\scase{B}(1)}$, whereas the remaining terms are canceled by the diamagnetic contribution. Therefore, $\boldsymbol{\sigma}^{\scase{C}(1)}=\boldsymbol{\sigma}^{\scase{B}(1)}$, and all four methods become equivalent.

Proceeding to the second-order response, hereafter for simplicity we limit our analysis to the second-harmonic generation (SHG) process, \ie $\omega_p=\omega_q\equiv\omega$. However, the conclusions are generally valid for other second-order processes. The SHG conductivities obtained in the LG, \ie $\boldsymbol{\sigma}^{\scase{A}(2)}$ and $\boldsymbol{\sigma}^{\scase{B}(2)}$, are related to each other via
\begin{align}
	\label{eq:ConnectAB}
	\boldsymbol{\sigma}^{\scase{B}(2)} = \boldsymbol{\sigma}^{\scase{A}(2)} + C_{ee} \sum_{nm} \dfrac{\va{X}_n \va{\Pi}_{nm} \va{X}_m^*}{(\hbar\omega+E_n)(\hbar\omega+E_m)} \, .
\end{align} 
This is seen by rewriting the frequency dependent terms in $\boldsymbol{\sigma}^{\scase{B}(2)}$, \eg $2\hbar\omega/(2\hbar\omega-E_n)=1+E_n/(2\hbar\omega-E_n)$.
The extra term on the right-hand side of Eq.~(\ref{eq:ConnectAB}) can be shown to vanish by exchanging the dummy indices, $m \leftrightarrow n$, and noticing that $\va{\Pi}_{nm}=\va{\Pi}^*_{mn}=-\va{\Pi}_{mn}$ due to the time-reversal symmetry (see Appendix~\ref{sec:AppendixB}). 
For the SHG conductivities in the VG, \ie $\boldsymbol{\sigma}^{\scase{C}(2)}$ and $\boldsymbol{\sigma}^{\scase{D}(2)}$, it is straightforward to show that 
\begin{subequations}
	\begin{align}
		\label{eq:ConnectCD}
		\boldsymbol{\sigma}^{\scase{D}(2)} &= \boldsymbol{\sigma}^{\scase{C}(2)}+ \dfrac{iC_{ee}}{(\hbar\omega)^2} \sum_{nm} \va{\Pi}_n \va{X}_{nm} \va{\Pi}_m^* \nonumber \\ 
		& \hspace{2.8cm} \times \dfrac{(2\hbar\omega+E_m-E_n)}{(\hbar\omega+E_n)(\hbar\omega+E_m)} \, , \\
		\label{eq:ConnectBD}
		\boldsymbol{\sigma}^{\scase{D}(2)} &= \boldsymbol{\sigma}^{\scase{B}(2)} + 2C_{ee} \sum_{nm} \dfrac{\va{X}_n \va{\Pi}_{nm} \va{X}_m^*}{(\hbar\omega+E_n)(\hbar\omega+E_m)} \, .
	\end{align}
\end{subequations} 
To derive these relations, the frequency-dependent fractions of Eqs.~(\ref{eq:TensorSecondC}) and (\ref{eq:TensorSecondD}) have been rewritten using the same technique as employed for deriving Eq.~(\ref{eq:ConnectAB}).
For Eq.~(\ref{eq:ConnectCD}), as well as Eq.~(\ref{eq:ConnectBD}), the second term on the right-hand side vanishes due to the time-reversal symmetry as in Eq.~(\ref{eq:ConnectAB}). So, despite the fact that $\boldsymbol{\sigma}^{\scase{A}(2)}$, $\boldsymbol{\sigma}^{\scase{B}(2)}$, $\boldsymbol{\sigma}^{\scase{C}(2)}$ and $\boldsymbol{\sigma}^{\scase{D}(2)}$ differ in form, they are equivalent regardless of the number of bands used in the calculation. In particular, we note that the zero-frequency divergences of $\boldsymbol{\sigma}^{\scase{C}(2)}$ and $\boldsymbol{\sigma}^{\scase{D}(2)}$ are only apparent.

Now, let us focus on the CVG, \ie tensors labeled by $\scase{C}'$. One can show that the conductivity tensors obtained by method $\scase{C}'$ using Eqs.~(\ref{eq:TensorFirstE}) and (\ref{eq:TensorSecondE}) include several additional non-vanishing terms compared to the other four methods. Here, we demonstrate this fact for the linear response function, Eq.~(\ref{eq:TensorFirstE}), but the same conclusion can be drawn for the quadratic response, Eq.~(\ref{eq:TensorSecondE}).
Using Eq.~(\ref{eq:CouplingMatrixA}), it is straightforward to show that
\begin{subequations}
	\label{eq:ConventionalVG}
	\begin{align}
		&\va{\Pi}_n = -iE_n\va{X}_n = \va{P}_n -i \va{F}_n \, , \\ 
		&\va{F}_n \equiv \sum_{cv\va{k}} \psi_{cv\va{k}}^{(n)} \va{f}_{vc\va{k}} \, , 
	\end{align}
\end{subequations}
where $\va{f}_{vc\va{k}} \equiv \sum_{c'v'\va{k}'} W_{c'v'\va{k}', cv\va{k}} \va{x}_{v'c'\va{k}'}$. The value of $\va{F}_n$ depends on the strength of the electron-hole interaction, and, hence, vanishes when excitonic effects are ignored.
Thus, if $\va{P}_n$ in Eq.~(\ref{eq:TensorFirstE}) is replaced by $\va{\Pi}_n+i\va{F}_n$, we obtain $\boldsymbol{\sigma}^{\scase{C}'(1)}=\boldsymbol{\sigma}^{\scase{C}(1)} + \textrm{``extra term"}$, where the ``extra term" depends on the value of $\va{F}_n$. We confirm numerically that this term is generally nonzero and contributes to the conductivity $\boldsymbol{\sigma}^{\scase{C}'(1)}$, which makes it different from the other four methods. Indeed, we will demonstrate numerically in Sec.~\ref{sec:Results} that by decreasing the effect of the Coulomb potential and including more bands in the calculation the ``extra term" contributes less and, hence, $\boldsymbol{\sigma}^{\scase{C}'(1)}$ converges toward $\boldsymbol{\sigma}^{\scase{A}(1)}$-$\boldsymbol{\sigma}^{\scase{D}(1)}$. The same behavior should follow for the nonlinear responses obtained using the CVG interaction. This is readily seen by noticing that $\va{\Pi}_{nm} = i(E_n-E_m)\va{X}_{nm} = \va{P}_{nm} +i \va{F}_{nm}$, where $\va{F}_{nm}$ has a complicated form written in terms of $W_{c'v'\va{k}', cv\va{k}}$, analogous to Eq.~
(\ref{eq:ConventionalVG}). Hence, it is straightforward to confirm that $\boldsymbol{\sigma}^{\scase{C}'(2)}=\boldsymbol{\sigma}^{\scase{C}(2)} + \textrm{``extra term"}$, where the non-vanishing ``extra term" here is a function of both $\va{F}_n$ and $\va{F}_{nm}$.


\section{Numerical Results \label{sec:Results}}
\begin{figure}[t]
	\includegraphics[width=0.49\textwidth]{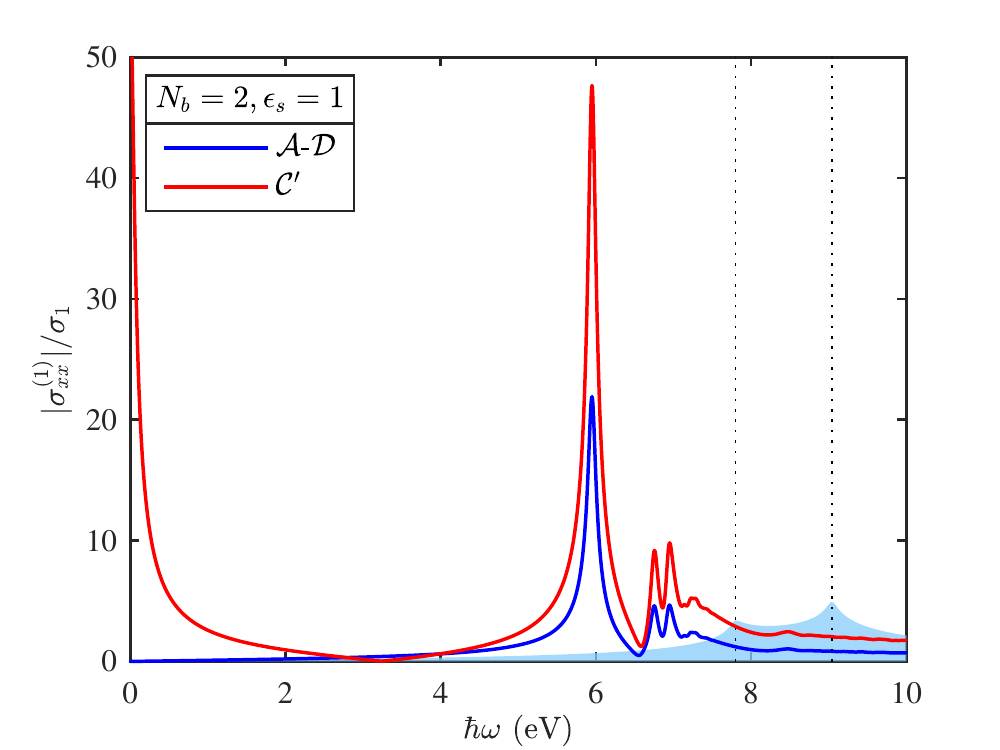}
	\caption[OC with epsilon1]{Excitonic OC spectrum of hBN monolayer obtained by methods $\scase{A}$-$\scase{D}$ (blue), and $\scase{C}'$ (red) for $N_b=2$. The values are normalized to $\sigma_1 \equiv e^2/4\hbar=\SI{6.0853e-5}{S}$. For comparison purposes, the OC spectrum without excitonic effects found in method $\scase{A}$ is also plotted (filled light blue). The black dotted lines indicate $\hbar\omega=\{E_g, E_\mathrm{vH}\}$. }
	\label{fig:Spectrum1st}
\end{figure}

\begin{figure}[t]
	\includegraphics[width=0.49\textwidth]{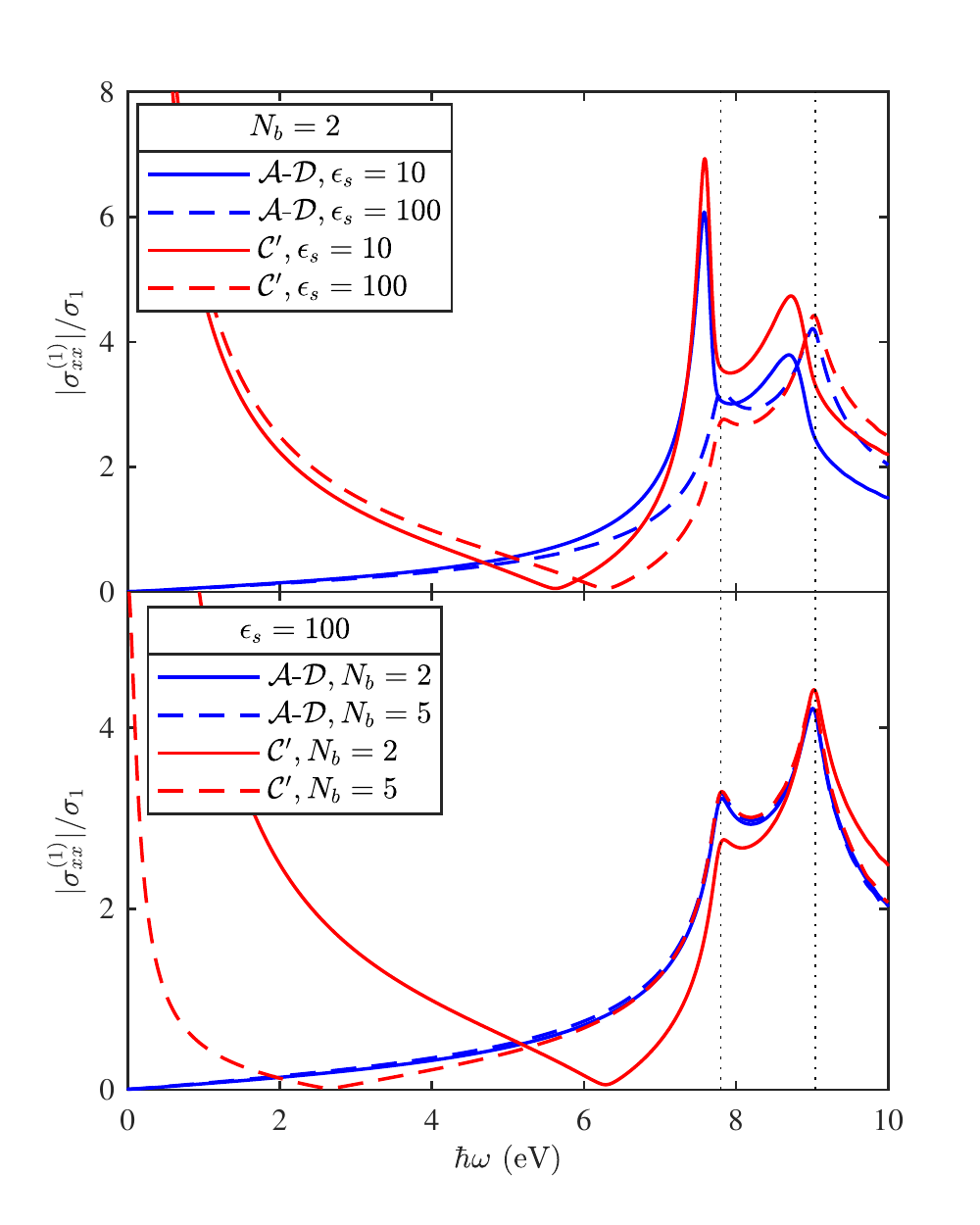}
	\caption[OC with large epsilon]{Excitonic OC spectrum of hBN monolayer obtained by methods $\scase{A}$-$\scase{D}$ (blue), and $\scase{C}'$ (red) for different values of $\epsilon_s$ and $N_b$. Top panel: $\epsilon_s=10$ (solid) and $\epsilon_s=100$ (dashed) for $N_b=2$. Bottom panel: $N_b=2$ (solid) and $N_b=5$ (dashed) for $\epsilon_s=100$. Here, the line shape broadening is set to $\eta=0.1$ eV, and the OC values are normalized to $\sigma_1$ (see Fig.~\ref{fig:Spectrum1st} caption). The black dotted lines mark $\hbar\omega=\{E_g, E_\mathrm{vH}\}$.  }
	\label{fig:Spectrum1st_Eps}
\end{figure}

\begin{figure}[t]
	\includegraphics[width=0.49\textwidth]{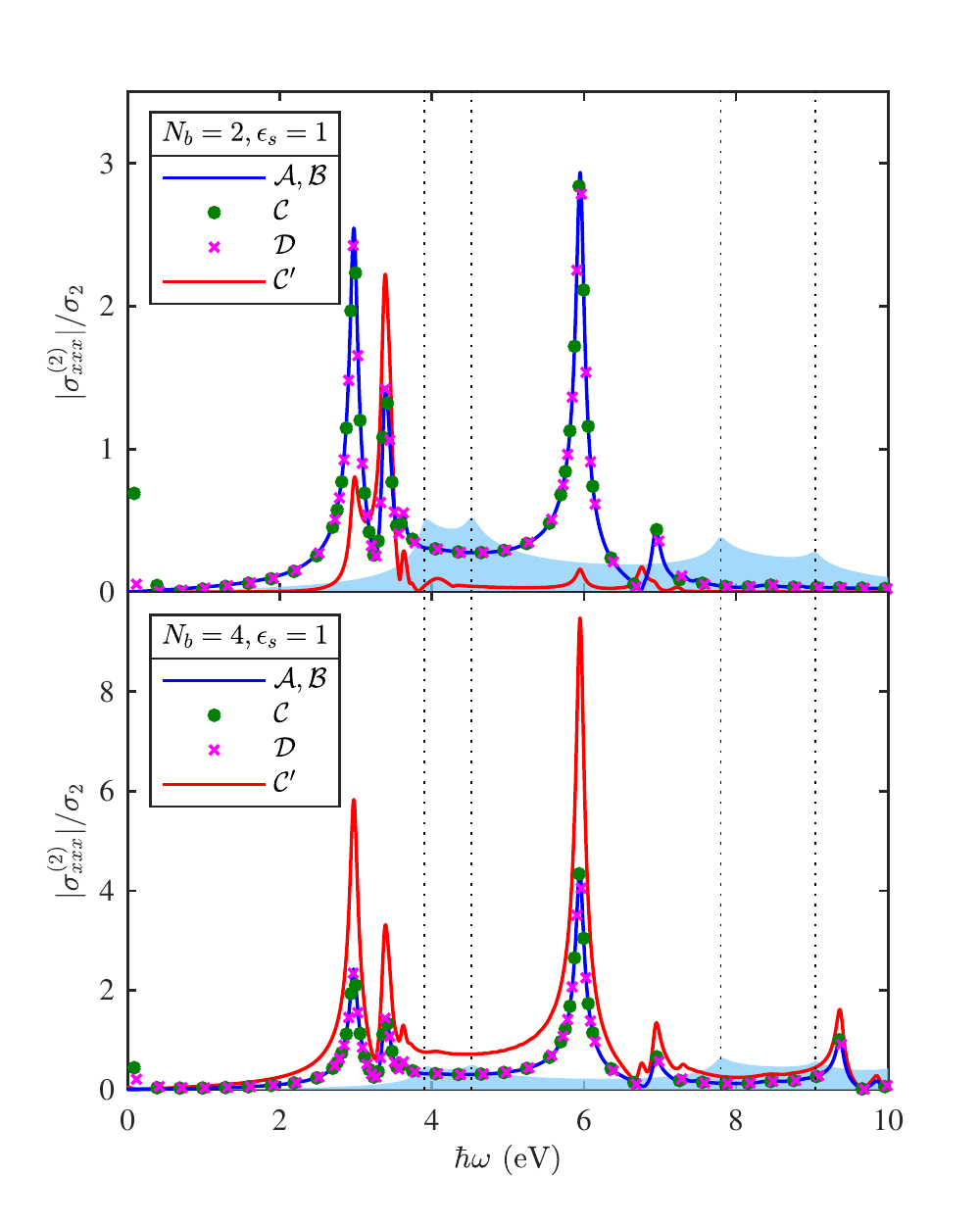}
	\caption[SHG spectra]{Excitonic SHG spectrum of hBN monolayer obtained from methods $\scase{A},\scase{B}$ (blue solid lines), $\scase{C}$ (green circles), $\scase{D}$ (magenta crosses), and $\scase{C}'$ (red solid lines) for $N_b=2$ (top panel) and $N_b=4$ (bottom panel). The values are normalized to $\sigma_2 \equiv e^3a/4\gamma_0\hbar=\SI{6.559e-15}{SmV^{-1}}$, where we set $\gamma_0=2.33$ eV. For comparison purposes, the SHG conductivity spectrum of method $\scase{A}$ in the IPA (filled light blue) is also shown. The dotted lines from the left to right indicate $2\hbar\omega=E_g$, $2\hbar\omega=E_\mathrm{vH}$, $\hbar\omega=E_g$, and  $\hbar\omega=E_\mathrm{vH}$, respectively.}
	\label{fig:Spectrum2nd}
\end{figure}

In this section, we apply the proposed theory to compute the excitonic optical response of hBN monolayers, and compare the calculated OC and SHG spectra generated by the five methods of Table~\ref{tb:Gauges}. The single-particle band structure and required matrix elements are obtained from an empirical pseudopotential Hamiltonian \cite{Yu2010}. This approach, which accurately reproduces the low-energy properties of hBN monolayers, allows us to have access to a large number of bands. The pseudopotential parameterization is reported in our previous work, see Ref.~\onlinecite{Taghizadeh2017}. For the present numerical examples, we have used 85 reciprocal lattice vectors in the pseudopotential implementation, which generates a total of 85 bands including one valence (the band with lowest energy) and 84 conduction bands. In our numerical implementation, we assume that the eigenenergies obtained by the pseudopotential correspond to the quasi-particle energies, and the pseudopotential wavefunctions are used for computing all matrix elements. For instance, the interband position matrix elements are obtained using $i(\varepsilon_{n\va{k}}-\varepsilon_{m\va{k}})\ev{n\va{k}|\hat{\va{x}}^{(e)}|m\va{k}}=\ev{n\va{k}|\hat{\va{p}}|m\va{k}}$, where $\varepsilon_{n\va{k}}$ and $|m\va{k}\rangle$ are the pseudopotential energies and wavefunctions, respectively. 
Out of the 85 available bands, only the $N_b \geq 2$ lowest bands are included in the calculations. To ensure a proper convergence of the results, more than 11000 $\va{k}$-points are used for discretizing the first BZ.
A lattice constant of $a=2.51\AA$ is assumed and the quasi-particle band gap and van Hove transition energies are $E_g=7.78$ eV and $E_\mathrm{vH}=9.04$ eV, respectively. Due to the symmetry of the honeycomb lattice in hBN monolayers, it is sufficient to study only the diagonal components of the conductivity tensors, \ie $\sigma_{xx}^{(1)}$ and $\sigma_{xxx}^{(2)}$ \cite{Hipolito2016}.
Finally, the line shape broadening is accounted for by adding a small phenomenological imaginary part, $i\eta$, to the frequency, \ie $\omega \rightarrow \omega+i\eta$. We set $\eta=0.05$ eV for Figs.~\ref{fig:Spectrum1st} and \ref{fig:Spectrum2nd}, whereas it is increased to $\eta=0.1$ eV for Fig.~\ref{fig:Spectrum1st_Eps} to ensure sufficiently smooth curves.

It is well-known that for a realistic description of the exciton spectrum of 2D materials, the Coulomb potential should be accurately screened. 
However, the screening is not properly included in the MFA and, hence, it is introduced phenomenologically \cite{Schafer2002}. In the present work, we use the Keldysh potential for the direct Coulomb interaction, which is a widely-accepted form of the screened potential for 2D materials \cite{Cudazzo2011, Qiu2016, Galvani2016, Trolle2017, Thygesen2017}. In real space, the Keldysh potential is given by
\begin{align}
	\mathcal{V}^d(\va{r}) &= C_0 \dfrac{\pi}{2r_0} \bigg[ \mathbb{H}_0\Big(\frac{\epsilon_s r}{r_0}\Big) - \mathbb{Y}_0\Big(\frac{\epsilon_s r}{r_0}\Big)\bigg] \, ,
\end{align}
where $C_0 \equiv e^2/4\pi\epsilon_0$, $r=|\va{r}|$, and $\mathbb{H}_0$ and $\mathbb{Y}_0$ are the Struve function and Bessel function of second type, respectively. The two parameters $\epsilon_s$ and $r_0$ are the substrate screening and screening length, respectively, which are set to $\epsilon_s=1$ and $r_0=10\AA$ for freely-suspended hBN monolayers \cite{Galvani2016}. The Fourier transform of the Keldysh potential reads $\mathcal{V}^d(\va{q}) \equiv C_0 2\pi /q(\epsilon_s+r_0q)$, which is used for obtaining the Coulomb matrix elements according to Eq.~(\ref{eq:CoulombMatrix}). The summation over $\va{G}$ in Eq.~(\ref{eq:CoulombMatrix}) is truncated to the seven smallest reciprocal vectors, since the impact of larger $\va{G}$'s becomes negligible. Regarding the exchange terms, we neglect them due to their minor impact on the results \cite{Thygesen2017}. 
We note that the screening of the Coulomb potential influences the shape of the spectrum, yet our conclusions concerning gauge invariance are independent of the screening model.

Figure~\ref{fig:Spectrum1st} shows $|\sigma_{xx}^{(1)}|$ of suspended hBN monolayers versus frequency obtained using Eqs.~(\ref{eq:TensorFirstA})-(\ref{eq:TensorFirstE}) for $N_b=2$. For comparison purposes, we also plot the OC computed in the IPA limit simply by increasing the screening, \ie $\epsilon_s \rightarrow \infty$. Without excitons, the response shows the expected features associated with the band gap and van Hove singularity \cite{Hipolito2016}. In contrast, including the excitonic effects dramatically changes the spectrum by introducing a strong peak below the band gap at approximately 5.95 eV due to the fundamental exciton, while several other strong peaks are formed due to higher-order excitons. The excitonic OC spectrum is in good qualitative agreement with the previous results for hBN monolayers in Refs.~\onlinecite{Wirtz2006, Attaccalite2011, Pedersen2015}.

Now, let us focus on the differences between the excitonic responses computed by the five methods. The results in Fig.~\ref{fig:Spectrum1st} confirm that the spectra generated by methods $\scase{A}$ to $\scase{D}$ are numerically identical, whereas the spectrum obtained by method $\scase{C}'$ is considerably different. For instance, $\boldsymbol{\sigma}^{\scase{C}'(1)}$ suffers from a zero-frequency divergence, in contrast to the divergence-free $\boldsymbol{\sigma}^{\scase{A}(1)}$ to $\boldsymbol{\sigma}^{\scase{D}(1)}$. In addition, method $\scase{C}'$ overestimates the magnitude of the response function substantially over the whole frequency range. For any finite Coulomb screening $\epsilon_s$, the differences between $\boldsymbol{\sigma}^{\scase{C}'(1)}$ and $\boldsymbol{\sigma}^{\scase{A}(1)}$-$\boldsymbol{\sigma}^{\scase{D}(1)}$ persist, and they do not disappear even for a complete basis set. Nonetheless, for a very large screening value, the disagreement between $\scase{C}'$ and $\scase{A}$-$\scase{D}$ diminishes by including more bands in the calculation as discussed in Sec.~\ref{sec:Gauge}. This is illustrated quantitatively in Fig.~\ref{fig:Spectrum1st_Eps}, where the OC spectra computed by methods $\scase{A}$-$\scase{D}$ and $\scase{C}'$ are displayed for two representative value of substrate screening, namely $\epsilon_s=\{10, 100\}$, with $N_b=2$ in the top panel. In the bottom panel of Fig.~\ref{fig:Spectrum1st_Eps}, we plot the same spectra for the larger screening value, \ie $\epsilon_s=100$, with $N_b=2$ and $N_b=5$. Increasing $N_b$ from 2 to 5 barely influences the response generated by methods $\scase{A}$-$\scase{D}$, whereas the results of method $\scase{C}'$ differ considerably. Furthermore, the OC obtained by the CVG, \ie $\boldsymbol{\sigma}^{\scase{C}'(1)}$, converges toward the results generated by other methods if both the screening and basis set size is increased, which is in agreement with the IPA results reported in Ref.~\onlinecite{Taghizadeh2017}.

Proceeding to the nonlinear response, Fig.~\ref{fig:Spectrum2nd} illustrates the SHG conductivities computed by methods $\scase{A}$-$\scase{D}$ and $\scase{C}'$ for two representative sizes of the basis set, $N_b=2$ in the top panel and $N_b=4$ in the bottom one. The SHG conductivities obtained in the IPA limit are also depicted for comparison. These responses agree with the results in Ref.~\onlinecite{Pedersen2015}. 
Beginning with the IPA result, the spectrum shows the features associated with $2\hbar\omega \sim \{E_g, E_\mathrm{vH}\}$ and $\hbar\omega \sim \{E_g, E_\mathrm{vH}\}$. Including more bands in the calculations barely changes the low-frequency resonances at $2\hbar\omega \sim \{E_g, E_\mathrm{vH}\}$, whereas it enhances the high-frequency resonances mainly due to an interband contribution caused by higher conduction bands \cite{Taghizadeh2017}. Adding excitons to the SHG response leads to a strong modification of the spectrum similar to the linear response, \eg several strong resonances are formed by excitons at frequencies below $\{E_g/2,E_g\}$ \cite{Gruning2014, Pedersen2015}. 


Focusing on the excitonic SHG responses, the results show that the LG conductivities are numerically identical for the both value of $N_b$, \ie $\sigma_{xxx}^{\scase{A}(2)}=\sigma_{xxx}^{\scase{B}(2)}$. In addition, the conductivities computed in the VG with the correct interaction Hamiltonian, \ie tensors labeled by $\scase{C}$ and $\scase{D}$, essentially agree with the calculations in the LG. The tiny differences between $\sigma_{xxx}^{\scase{C}(2)}$/$\sigma_{xxx}^{\scase{D}(2)}$ and $\sigma_{xxx}^{\scase{A}(2)}$/$\sigma_{xxx}^{\scase{B}(2)}$ at low frequencies are mainly due to the BZ discretization, and diminish by using a finer $\va{k}$-mesh. In particular, the zero-frequency divergences of methods $\scase{C}$ and $\scase{D}$ are only apparent. In contrast, the SHG responses found by method $\scase{C}'$ do not agree with the other four methods for both values of $N_b$. In particular, $\sigma_{xxx}^{\scase{C}'(2)}$ varies dramatically when more bands are included in the calculations, and the result for $N_b=2$ is highly inaccurate. In addition, even for $N_b=4$, method $\scase{C}'$ overestimates the magnitudes of both $\omega$ and $2\omega$ resonances by roughly a factor of two. Nonetheless, method $\scase{C}'$ converges toward the conductivities computed by the other methods if $\epsilon_s \rightarrow \infty$ and $N_b \rightarrow \infty$, similarly to the OC responses discussed before. Summarizing, the excitonic SHG conductivities obtained using Eqs.~(\ref{eq:TensorSecondA})-(\ref{eq:TensorSecondD}) are equivalent regardless of the number of bands in the calculations, whereas the SHG response computed by the CVG, \ie Eq.~(\ref{eq:TensorSecondE}), does not agree with the rest even for a complete basis set.


\section{Summary \label{sec:Conclusion}}
In summary, we have theoretically investigated the gauge invariance of linear and nonlinear optical responses when excitonic effects are included. The expressions for conductivity tensors were derived rigorously in the density matrix framework within the MFA for a multiband semiconductor. We have considered four distinct theoretical approaches derived from the combination of two choices of gauge and two ways of evaluating the current density, \ie directly or via the polarization. We have shown both analytically and numerically that by using the correct interaction Hamiltonian and observable in the VG, both the linear and quadratic responses obtained by the four methods become identical. The correct interaction in VG should be written in terms of the Heisenberg momentum $\hat{\va{\Pi}}$, defined as the commutator of the unperturbed Hamiltonian and position operators. Despite the equivalence of the four methods, computing the conductivities in the LG, \ie tensors labeled with $\scase{A}$ and $\scase{B}$, is more straightforward than the VG, \ie tensors labeled with $\scase{C}$ and $\scase{D}$. Finally, the excitonic optical responses generated by the CVG interaction, \ie tensors labeled with $\scase{C}'$, do not agree with the other methods, since $\hat{\va{\Pi}}$ is generally different from the momentum operator $\hat{\va{P}}$ when electron-hole interaction is included. The present formalism can readily be extended to generate gauge invariant responses for higher-order nonlinear processes.

\acknowledgments
The authors thank F. Hipolito, J. Have, and F. Bonabi for helpful discussions throughout the project. This work was supported by the QUSCOPE center sponsored by the Villum Foundation and TGP is financially supported by the CNG center under the Danish National Research Foundation, project DNRF103.

\appendix

\begin{widetext}
\section{Equation of motion and its perturbative solution \label{sec:AppendixA}}
Here, we review the derivation of the dynamical equation for the density matrix, and present its perturbative solution up to the second order. Our starting point is the many-body Hamiltonian in second quantization, Eq.~(\ref{eq:TotalHamiltonian}). This Hamiltonian leads to the usual equation of motion (quantum Liouville) for the density matrix $\rho_{ji}$ \cite{Pedersen2015},
\begin{align}
	\label{eq:DensityGeneral}
	i\hbar \pdv{\rho_{ji}}{t} - \varepsilon_{ji} \rho_{ji} &- \sum_{lmn} \big(\mathcal{V}_{mlni} -\mathcal{V}_{lmni} \big) \big(\rho_{nl}-\delta_{mi}\delta_{nl}\delta_{lv}\big) \rho_{jm} \nonumber \\
	&- \sum_{lmn} \big(\mathcal{V}_{jlmn} -\mathcal{V}_{jlnm} \big) \big(\rho_{nl}-\delta_{mj}\delta_{nl}\delta_{lv}\big) \rho_{mi} = \sum_{l} \big(u_{jl}\rho_{li} - u_{li}\rho_{jl} \big) \, ,
\end{align}
where $\varepsilon_{ji} \equiv \varepsilon_j-\varepsilon_i$ and the quasi-particles energies $\varepsilon_i \equiv \varepsilon_i^0+\sum_l (\mathcal{V}_{ilil}-\mathcal{V}_{illi})\delta_{lv}$ are introduced, with the Kronecker delta serving to count occupied states only.

For the special case of Bloch states, each index should run over both band index and wavevector. To proceed, we assume that the density matrix is diagonal with respect to the wavevector, \ie $\rho_{j\va{k}_j i\va{k}_i} \equiv \rho_{ji\va{k}_i} \delta_{\va{k}_i,\va{k}_j}$ \cite{Pedersen2015}, since the diagonal part of density matrix is the dominant contribution to the system response. 
Hence, the dynamical equation for the density matrix in crystals reads
\begin{subequations}
	\label{eq:DensityInit}
	\begin{align}
		i\hbar \pdv{\rho_{ji\va{k}}}{t} -& \varepsilon_{ji\va{k}} \rho_{ji\va{k}} - \dfrac{1}{A} \sum_{lmn\va{k}'} \Big[ \mathcal{V}_{mlni}^d(\va{k},\va{k}') - g\mathcal{V}_{lmni}^x(\va{k}',\va{k}) \Big] \big(\rho_{nl\va{k}'}-\delta_{mi}\delta_{nl}\delta_{lv}\big) \rho_{jm\va{k}} \nonumber \\
		&- \dfrac{1}{A} \sum_{lmn\va{k}'} \Big[ g\mathcal{V}_{jlmn}^x(\va{k},\va{k}') - \mathcal{V}_{jlnm}^d(\va{k},\va{k}') \Big] \big(\rho_{nl\va{k}'}-\delta_{mj}\delta_{nl}\delta_{lv}\big) \rho_{mi\va{k}} = \sum_{l} \big(u_{jl\va{k}}\rho_{li\va{k}} - u_{li\va{k}}\rho_{jl\va{k}} \big) \, ,
	\end{align}
\end{subequations}
where $\varepsilon_{ji\va{k}} \equiv \varepsilon_{j\va{k}}-\varepsilon_{i\va{k}}$, and the extra factors of $g$ appear due to the spin degeneracy of singlet states \cite{Leung1997}. The direct and exchange Coulomb matrix elements $\mathcal{V}_{abcd}^d$ and $\mathcal{V}_{abcd}^x$ read
\begin{subequations}
	\label{eq:CoulombMatrix}
	\begin{align}
		&\mathcal{V}_{abcd}^d(\va{k},\va{k}') = \sum_{\va{G}} I_{a\va{k},c\va{k}'}(\va{G}) I_{b\va{k}',d\va{k}}(-\va{G}) \mathcal{V}^d(\va{k}-\va{k}'-\va{G}) \, , \\
		&\mathcal{V}_{abcd}^x(\va{k},\va{k}') = \sum_{\va{G} \neq \va{0}} I_{a\va{k},c\va{k}}(\va{G}) I_{b\va{k}',d\va{k}'}(-\va{G}) \mathcal{V}^x(-\va{G}) \, .
	\end{align}
\end{subequations}
Here, the summation is performed over reciprocal vectors $\va{G}$, and the Bloch overlaps $I_{a\va{k},c\va{k}'}(\va{G}) \equiv A_\textrm{uc}^{-1} \int_\textrm{uc} \varphi_{a\va{k}}^*(\va{r}) \varphi_{c\va{k}'}(\va{r}) \exp(i\va{G} \vdot \va{r}) \dd[D]{\va{r}}$ are introduced. In Eq.~(\ref{eq:CoulombMatrix}), $\mathcal{V}^d$ and $\mathcal{V}^x$ on the right-hand side are the Fourier transforms of the direct and exchange Coulomb potential, respectively. Note that the long range contribution of the exchange part, \ie $\va{G} = \va{0}$, is removed \cite{Onida2002, Leitsmann2005, Koskelo2017}. 
The equation of motion for $\rho_{ji\va{k}}(t)$, Eq.~(\ref{eq:DensityInit}), is solved perturbatively by iteration to any order of perturbation, \ie $\rho_{ji\va{k}}(t)=\sum_N \rho_{ji\va{k}}^{(N)}(t)$. The unperturbed solution, \ie  $\rho_{ji\va{k}}^{(0)}$, for the case of cold clean semiconductors is given by $\rho_{vv'\va{k}}^{(0)} = \delta_{vv'}$ and $\rho_{cc'\va{k}}^{(0)} = \rho_{cv\va{k}}^{(0)} = \rho_{vc\va{k}}^{(0)} = 0$. To the first order, $\rho_{cc'\va{k}}^{(1)} \approx 0$ and $\rho_{vv'\va{k}}^{(1)} \approx 0$, \ie the field-induced changes in the band occupation is negligible \cite{Pedersen2015}. 
Furthermore, the equation of motion for $\rho_{cv\va{k}}^{(1)}$ reads
\begin{align}
	\label{eq:DensityFirst}
	&i\hbar \pdv{\rho_{cv\va{k}}^{(1)}}{t} - \sum_{c'v'\va{k'}} H_{cv\va{k},c'v'\va{k}'} \rho_{c'v'\va{k}'}^{(1)} - \sum_{c'v'\va{k'}} T_{cv\va{k},c'v'\va{k}'} \rho_{v'c'\va{k}'}^{(1)} = u_{cv\va{k}}^{(1)}(t)  \, , 
\end{align}
where $u_{cv\va{k}}^{(1)}(t)$ is the first-order contribution of the perturbation, and $H_{cv\va{k},c'v'\va{k}'}$ and $T_{cv\va{k},c'v'\va{k}'}$ are defined as
\begin{subequations}
	\label{eq:CouplingMatrix}
	\begin{align}
		\label{eq:CouplingMatrixA}
		&H_{cv\va{k},c'v'\va{k}'} \equiv \varepsilon_{cv\va{k}} \delta_{c,c'}\delta_{v,v'}\delta_{\va{k},\va{k}'} + \dfrac{1}{A} [g\mathcal{V}_{cv'vc'}^x(\va{k},\va{k}')-\mathcal{V}_{cv'c'v}^d(\va{k},\va{k}')] \equiv \varepsilon_{cv\va{k}} \delta_{c,c'}\delta_{v,v'}\delta_{\va{k},\va{k}'} + W_{cv\va{k},c'v'\va{k}'} \, , \\
		\label{eq:CouplingMatrixB}
		&T_{cv\va{k},c'v'\va{k}'} \equiv \dfrac{1}{A} [g\mathcal{V}_{cc'vv'}^x(\va{k},\va{k}')-\mathcal{V}_{cc'v'v}^d(\va{k},\va{k}')] \, .
	\end{align}
\end{subequations}
Similarly, the equation of motion for $\rho_{vc\va{k}}^{(1)}$ is found by taking the complex conjugate of Eq.~(\ref{eq:DensityFirst}). 
One may solve the full coupled set of equations for $\rho_{cv\va{k}}^{(1)}$ and $\rho_{vc\va{k}}^{(1)}$. However, the $T_{cv\va{k},c'v'\va{k}'}$ terms can be ignored due to their small magnitude when compared to $H_{cv\va{k},c'v'\va{k}'}$,  because $|I_{c\va{k},v\va{k}'}| \ll |I_{c\va{k},c'\va{k}'}| , |I_{v\va{k},v'\va{k}'}|$. This leads to the decoupling of $\rho_{cv\va{k}}^{(1)}$ and $\rho_{vc\va{k}}^{(1)}$ equations, which is known as the Tamm-Dancoff approximation \cite{Benedict1998, Rohlfing2000}. 
Going one step further, the dynamical equations for the second-order density matrix in the Tamm-Dancoff approximation read
\begin{subequations}
	\label{eq:DensitySecond}
	\begin{align}
		\label{eq:DensitySecondA}
		&i\hbar \pdv{\rho_{cv\va{k}}^{(2)}}{t} - \sum_{c'v'\va{k'}} H_{cv\va{k},c'v'\va{k}'} \rho_{c'v'\va{k}'}^{(2)} = \sum_{c'} u_{cc'\va{k}}^{(1)}(t) \rho_{c'v\va{k}}^{(1)} - \sum_{v'} u_{v'v\va{k}}^{(1)}(t) \rho_{cv'\va{k}}^{(1)} + u_{cv\va{k}}^{(2)}(t) \, , \\ 
		&i\hbar \pdv{\rho_{cc'\va{k}}^{(2)}}{t} - \varepsilon_{cc'\va{k}} \rho_{cc'\va{k}}^{(2)} - \sum_{v'}\sum_{c_1 v_1\va{k}_1} W_{c'v'\va{k},c_1v_1\va{k}_1} \rho_{v_1c_1\va{k}_1}^{(1)} \rho_{cv'\va{k}}^{(1)} + \sum_{v'}\sum_{c_1 v_1\va{k}_1} W_{cv'\va{k},c_1v_1\va{k}_1} \rho_{c_1v_1\va{k}_1}^{(1)} \rho_{v'c'\va{k}}^{(1)} \nonumber \\
		&\hspace{7cm} = \sum_{v'} u_{cv'\va{k}}^{(1)}(t) \rho_{v'c'\va{k}}^{(1)} - \sum_{v'} u_{v'c\va{k}}^{(1)}(t) \rho_{cv'\va{k}}^{(1)} \, , \\ 
		&i\hbar \pdv{\rho_{vv'\va{k}}^{(2)}}{t} - \varepsilon_{vv'\va{k}} \rho_{vv'\va{k}}^{(2)} - \sum_{c'}\sum_{c_1 v_1\va{k}_1} W_{c'v'\va{k},c_1v_1\va{k}_1} \rho_{c_1v_1\va{k}_1}^{(1)} \rho_{v'c'\va{k}}^{(1)} + \sum_{c'}\sum_{c_1 v_1\va{k}_1} W_{c'v\va{k},c_1v_1\va{k}_1} \rho_{v_1c_1\va{k}_1}^{(1)} \rho_{c'v'\va{k}}^{(1)} \nonumber \\
		&\hspace{7cm} = \sum_{c'} u_{vc'\va{k}}^{(1)}(t) \rho_{c'v'\va{k}}^{(1)} - \sum_{c'} u_{c'v'\va{k}}^{(1)}(t) \rho_{vc'\va{k}}^{(1)} \, ,
	\end{align}
\end{subequations}
where $u_{cv\va{k}}^{(2)}(t)$ is the second-order contribution to the perturbation.
Similarly, the equation of motion for $\rho_{vc\va{k}}^{(2)}(t)$ is obtained by taking the complex conjugate of Eq.~(\ref{eq:DensitySecondA}).


The set of non-homogeneous equations of motion for the density matrix, Eqs.~(\ref{eq:DensityFirst}) and (\ref{eq:DensitySecond}), can be solved by employing Green's functions as explained in Ref.~\onlinecite{Pedersen2015}. This is done by diagonalizing the matrix $H_{eh}$ given in Eq.(\ref{eq:CouplingMatrixA}), \ie $H_{eh}|\psi^{(n)}\rangle=E_n|\psi^{(n)}\rangle$, which is essentially the well-known BSE.
Here, $E_n$ and $|\psi^{(n)}\rangle$ are the exciton energies and eigenstates, which are written in the basis of vertical transitions from valence to conduction bands, \ie $|\psi^{(n)}\rangle=\sum_{cv\va{k}} \psi_{cv\va{k}}^{(n)} |v\va{k}{\rightarrow}c\va{k}{\rangle}$. To continue, we consider an interaction potential of the form $\hat{U}(t) \equiv \hat{U}^{(1)} S(t) + \hat{U}^{(2)} S^2(t)$, where $S(t)$ is given as a set of time-harmonic terms, $S(t) \equiv 1/2\sum_{p} S(\omega_p) e^{-i\omega_p t}$.
Thus, the solutions of Eqs.~(\ref{eq:DensityFirst}) and (\ref{eq:DensitySecond}) read
\begin{subequations}
	\label{eq:DensityFinal}
	\begin{align}
		\label{eq:DensityFinalA}
		&\rho_{vv'\va{k}}^{(1)}(t) \approx 0 \, , \quad \rho_{cc'\va{k}}^{(1)}(t) \approx 0 \, , \\
		\label{eq:DensityFinalB}
		&\rho_{cv\va{k}}^{(1)}(t) = \dfrac{1}{2} \sum_p S(\omega_p) e^{-i\omega_p t} \Bigg\{ \sum_n \dfrac{\psi_{cv\va{k}}^{(n)} U_n^*}{\hbar \omega_p -E_n} \Bigg\}  = \rho_{vc\va{k}}^{(1)*}(t) \, ,  \\
		\label{eq:DensityFinalC}
		&\rho_{cv\va{k}}^{(2)}(t) = \dfrac{1}{4} \sum_{pq} S(\omega_p) S(\omega_q) e^{-i\omega_2t} \Bigg\{ \sum_{nm} \bigg[ \dfrac{\psi_{cv\va{k}}^{(n)} U_{nm} U_m^*}{(\hbar\omega_2-E_n)(\hbar\omega_q-E_m)} \bigg] + \sum_n \bigg[ \dfrac{\psi_{cv\va{k}}^{(n)} \bar{U}_n^*}{\hbar \omega_2 -E_n} \bigg] \Bigg\} = \rho_{vc\va{k}}^{(2)*}(t) \, , \\
		\label{eq:DensityFinalD}
		&\rho_{cc'\va{k}}^{(2)}(t) = -\dfrac{1}{4} \sum_{pq} S(\omega_p) S(\omega_q) e^{-i\omega_2t} \Bigg\{ \sum_{nm}  \dfrac{U_n U_m^*}{(\hbar\omega_q+E_n)(\hbar\omega_p-E_m)} \sum_{v_1} \psi_{c'v_1\va{k}}^{(n)*} \psi_{cv_1\va{k}}^{(m)} \Bigg\} \, , \\
		\label{eq:DensityFinalE}
		&\rho_{vv'\va{k}}^{(2)}(t) = \dfrac{1}{4} \sum_{pq} S(\omega_p) S(\omega_q) e^{-i\omega_2t} \Bigg\{ \sum_{nm} \dfrac{U_n U_m^*}{(\hbar\omega_q+E_n)(\hbar\omega_p-E_m)} \sum_{c_1} \psi_{c_1v\va{k}}^{(n)*} \psi_{c_1v'\va{k}}^{(m)} \Bigg\} \, , 
	\end{align}
\end{subequations}
where $\omega_2 \equiv \omega_p+\omega_q$ and the excitonic matrix elements of the perturbation are defined as:
\begin{align}
	\label{eq:MatrixElMB}
	&U_n \equiv \sum_{cv\va{k}} \psi_{cv\va{k}}^{(n)} u_{vc\va{k}}^{(1)} \, , \quad \bar{U}_n \equiv \sum_{cv\va{k}} \psi_{cv\va{k}}^{(n)} u_{vc\va{k}}^{(2)} \, , \quad U_{nm} \equiv \sum_{cv\va{k}} \psi_{cv\va{k}}^{(n)*} \bigg[ \sum_{c_1} \psi_{c_1v\va{k}}^{(m)} u_{cc_1\va{k}}^{(1)} - \sum_{v_1} \psi_{cv_1\va{k}}^{(m)} u_{v_1v\va{k}}^{(1)} \bigg] \, . 
\end{align}
Note that $U_n$, $\bar{U}_n$ are the matrix elements between the ground state and exciton eigenstates, \ie $U_n=\ev{\psi_0|\hat{U}^{(1)}|\psi^{(n)}}$, $\bar{U}_n=\ev{\psi_0|\hat{U}^{(2)}|\psi^{(n)}}$, and $U_{nm}$ corresponds to a matrix element between two exciton eigenstates, \ie $U_{nm}=\ev{\psi^{(n)}|\hat{U}^{(1)}|\psi^{(m)}}$. The second-order density matrix oscillates at frequency $\omega_2$, which describes various second-order processes such as SHG ($\omega_p=\omega_q$) or optical rectification ($\omega_p=-\omega_q$).

Upon obtaining the density matrix, the expectation value of any one-body operator, \ie an operator that acts on individual electrons, is determined straightforwardly. In second quantization, a one-body operator is given by $\hat{O}=\sum_{kl} o_{kl} \hat{c}_k^\dagger \hat{c}_l$, and its expectation value reads $\ev{\hat{O}}=\sum_{kl} o_{kl} \rho_{lk} = \textrm{tr}\{\hat{o}\hat{\rho}\}$. The operator $\hat{O}$ is assumed to contain a time-independent part, $\hat{O}^{(0)}$, and a part that is first-order in the perturbative field, $\hat{O}^{(1)}S(t)$, \ie $\hat{O} \equiv \hat{O}^{(0)}+\hat{O}^{(1)}S(t)$. Thus, the first- and second-order macroscopic responses of a system measured by $\hat{O}$ read
\begin{subequations}
	\label{eq:ObservableFinal}
	\begin{align}
		\label{eq:ObservableFinalA}
		&O^{(1)}(t) \equiv \tr{\hat{O}^{(0)}\rho^{(1)}} + \tr{\hat{O}^{(1)}\rho^{(0)}} = \dfrac{1}{2} \sum_p S(\omega_p) e^{-i\omega_p t} \Bigg \{ \sum_n \bigg[ \dfrac{O_n U_n^*}{\hbar\omega_p-E_n} - \dfrac{O_n^* U_n}{\hbar\omega_p+E_n} \bigg] + \sum_{v\va{k}} o_{vv\va{k}}^{(1)} \Bigg\} \, , \\
		\label{eq:ObservableFinalB}
		&O^{(2)}(t) \equiv \tr{\hat{O}^{(0)}\rho^{(2)}} + \tr{\hat{O}^{(1)}\rho^{(1)}} = \dfrac{1}{4} \sum_{pq} S(\omega_p) S(\omega_q) e^{-i\omega_2 t} \Bigg\{ \sum_{nm} \bigg[ \dfrac{O_n U_{nm} U_m^*}{(\hbar\omega_2-E_n)(\hbar\omega_q-E_m)}+ \nonumber \\
		&\dfrac{O_n^* U_{nm}^* U_m}{(\hbar\omega_2+E_n)(\hbar\omega_q+E_m)} - \dfrac{U_n O_{nm} U_m^*}{(\hbar\omega_q+E_n)(\hbar\omega_p-E_m)} \bigg] + \sum_n \bigg[ \dfrac{O_n \bar{U}_n^*}{\hbar\omega_2-E_n} - \dfrac{O_n^* \bar{U}_n}{\hbar\omega_2+E_n} \bigg] + \sum_n \bigg[ \dfrac{\bar{O}_n U_n^*}{\hbar\omega_p-E_n} - \dfrac{\bar{O}_n^* U_n}{\hbar\omega_p+E_n} \bigg] \Bigg\}  \, ,
	\end{align}
\end{subequations}
where the matrix elements of many-body observables $O_n$, $\bar{O}_n$ and $O_{nm}$ are defined analogous to their interaction counterpart, Eq.~(\ref{eq:MatrixElMB}), so that
\begin{align}
	&O_n \equiv \sum_{cv\va{k}} \psi_{cv\va{k}}^{(n)} o_{vc\va{k}}^{(0)} \, , \quad \bar{O}_n \equiv \sum_{cv\va{k}} \psi_{cv\va{k}}^{(n)} o_{vc\va{k}}^{(1)} \, , \quad O_{nm} \equiv \sum_{cv\va{k}} \psi_{cv\va{k}}^{(n)*} \bigg[ \sum_{c_1} \psi_{c_1v\va{k}}^{(m)} o_{cc_1\va{k}}^{(0)} - \sum_{v_1} \psi_{cv_1\va{k}}^{(m)} o_{v_1v\va{k}}^{(0)} \bigg] \, . 
\end{align}
We note that the last term in Eq.~(\ref{eq:ObservableFinalA}) is the matrix element of $\hat{O}^{(1)}$ with respect to the ground state, \ie $\sum_{v\va{k}} o_{vv\va{k}}^{(1)}=\ev{\psi_0|\hat{O}^{(1)}|\psi_0}$.

\section{Quadratic optical response \label{sec:AppendixB}}
The expressions for the second-order conductivities of the five methods in Table~\ref{tb:Gauges} are derived using Eq.~(\ref{eq:ObservableFinalB}), and given by
\begin{subequations}
	\label{eq:TensorSecond}
	\begin{align}
		\label{eq:TensorSecondA}
		\boldsymbol{\sigma}^{\scase{A}(2)} = 
		&-C_{ee} \sum_{nm} \bigg[\dfrac{\va{\Pi}_n \va{X}_{nm} \va{X}_m^*}{(\hbar\omega_2-E_n)(\hbar\omega_q-E_m)} + \dfrac{\va{\Pi}_n^* \va{X}_{nm}^* \va{X}_m}{(\hbar\omega_2+E_n)(\hbar\omega_q+E_m)} - \dfrac{\va{X}_n \va{\Pi}_{nm} \va{X}_m^*}{(\hbar\omega_q+E_n)(\hbar\omega_p-E_m)} \bigg]  \, , \\
		\label{eq:TensorSecondB}
		\boldsymbol{\sigma}^{\scase{B}(2)} = 
		&+C_{ee} (i\hbar\omega_2) \sum_{nm} \bigg[\dfrac{\va{X}_n \va{X}_{nm} \va{X}_m^*}{(\hbar\omega_2-E_n)(\hbar\omega_q-E_m)} + \dfrac{\va{X}_n^* \va{X}_{nm}^* \va{X}_m}{(\hbar\omega_2+E_n)(\hbar\omega_q+E_m)} - \dfrac{\va{X}_n \va{X}_{nm} \va{X}_m^*}{(\hbar\omega_q+E_n)(\hbar\omega_p-E_m)} \bigg] \, , \\
		\label{eq:TensorSecondC}
		\boldsymbol{\sigma}^{\scase{C}(2)} = &+\dfrac{C_{ee}}{(\hbar\omega_p)(\hbar\omega_q)} \sum_{nm} \bigg[ \dfrac{\va{\Pi}_n \va{\Pi}_{nm} \va{\Pi}_m^*}{(\hbar\omega_2-E_n)(\hbar\omega_q-E_m)} + \dfrac{\va{\Pi}_n^* \va{\Pi}_{nm}^* \va{\Pi}_m}{(\hbar\omega_2+E_n)(\hbar\omega_q+E_m)} - \dfrac{\va{\Pi}_n \va{\Pi}_{nm} \va{\Pi}_m^*}{(\hbar\omega_q+E_n)(\hbar\omega_p-E_m)} \bigg] \nonumber \\
		& -\dfrac{C_{ee}}{2i(\hbar\omega_p)(\hbar\omega_q)} \sum_{n} \bigg[ \dfrac{\va{\Pi}_n \va{A}_n^*}{\hbar\omega_2-E_n} + \dfrac{\va{\Pi}_n^*\va{A}_n }{\hbar\omega_2+E_n} \bigg] + \dfrac{C_{ee}}{i(\hbar\omega_p)(\hbar\omega_q)} \sum_{n} \bigg[ \dfrac{\va{A}_n\va{\Pi}_n^*}{\hbar\omega_p-E_n} + \dfrac{\va{A}_n^* \va{\Pi}_n}{\hbar\omega_p+E_n} \bigg] \, , \\
		\label{eq:TensorSecondD}
		\boldsymbol{\sigma}^{\scase{D}(2)} = 
		&-\dfrac{C_{ee}(i\hbar\omega_2)}{(\hbar\omega_p)(\hbar\omega_q)} \sum_{nm} \bigg[\dfrac{\va{X}_n \va{\Pi}_{nm} \va{\Pi}_m^*}{(\hbar\omega_2-E_n)(\hbar\omega_q-E_m)} + \dfrac{\va{X}_n^* \va{\Pi}_{nm}^* \va{\Pi}_m}{(\hbar\omega_2+E_n)(\hbar\omega_q+E_m)} - \dfrac{\va{\Pi}_n \va{X}_{nm} \va{\Pi}_m^*}{(\hbar\omega_q+E_n)(\hbar\omega_p-E_m)} \bigg] \nonumber \\
		&+ \dfrac{C_{ee}(\hbar\omega_2)}{2(\hbar\omega_p)(\hbar\omega_q)} \sum_{n} \bigg[ \dfrac{\va{X}_n\va{A}_n^*}{\hbar\omega_2-E_n} + \dfrac{\va{X}_n^*\va{A}_n}{\hbar\omega_2+E_n} \bigg] \, , \\
		\label{eq:TensorSecondE}
		\boldsymbol{\sigma}^{\scase{C}(2)} = &+\dfrac{C_{ee}}{(\hbar\omega_p)(\hbar\omega_q)} \sum_{n,m} \bigg[\dfrac{\va{P}_n \va{P}_{nm} \va{P}_m^*}{(\hbar\omega_2-E_n)(\hbar\omega_q-E_m)} + \dfrac{\va{P}_n^* \va{P}_{nm}^* \va{P}_m}{(\hbar\omega_2+E_n)(\hbar\omega_q+E_m)} - \dfrac{\va{P}_n \va{P}_{nm} \va{P}_m^*}{(\hbar\omega_q+E_n)(\hbar\omega_p-E_m)} \bigg]  \, ,
	\end{align}
\end{subequations}
where $C_{ee} \equiv ge^3\hbar^2/(m^3 A)$, $\va{A}_n \equiv \ev{\psi_0|[\hat{\va{X}},\hat{\va{\Pi}}]|\psi^{(n)}} = \sum_m (\va{X}_m\va{\Pi}_{mn} - \va{\Pi}_m\va{X}_{mn})$, $\va{\Pi}_{nm}=i(E_n-E_m)\va{X}_{nm}$, and $\va{P}_{nm}$ and $\va{X}_{nm}$ are defined using Eq.~(\ref{eq:MatrixElMB}) as
\begin{subequations}
	\begin{align}
		&\va{P}_{nm} \equiv \sum_{cv\va{k}} \psi_{cv\va{k}}^{(n)*} \bigg[ \sum_{c_1} \psi_{c_1v\va{k}}^{(m)} \va{p}_{cc_1\va{k}} - \sum_{v_1} \psi_{cv_1\va{k}}^{(m)} \va{p}_{v_1v\va{k}} \bigg] \, , \\
		\label{eq:PositionMatEl}
		&\va{X}_{nm} \equiv \sum_{cv\va{k}} \psi_{cv\va{k}}^{(n)*} \bigg[ \sum_{c_1} \psi_{c_1v\va{k}}^{(m)} \va{x}_{cc_1\va{k}} - \sum_{v_1} \psi_{cv_1\va{k}}^{(m)} \va{x}_{v_1v\va{k}} \bigg] \, . 
	\end{align}
\end{subequations}
Evaluating the momentum matrix elements $\va{P}_{nm}$ is rather straightforward, whereas the matrix elements of the ill-defined position operator in $\va{X}_{nm}$ should be separated to its interband and intraband parts as outlined in Sec.~\ref{sec:Optical}. 
Hence, we split the summations in this expression into two distinct contributions: $\va{X}_{nm} = \va{Y}_{nm} + m\va{Q}_{nm}/\hbar$, where $\va{Y}_{nm}$ and $\va{Q}_{nm}$ contain the interband ($c \neq c_1$ and $v \neq v_1$) and intraband ($c=c_1$ and $v=v_1$) components, respectively. So, $\va{Y}_{nm}$ and $\va{Q}_{nm}$ are given by
\begin{subequations}
	\begin{align}
		\label{eq:InterbandPosition}
		&\va{Y}_{nm} \equiv \sum_{cv\va{k}} \psi_{cv\va{k}}^{(n)*} \bigg[ \sum_{c_1\neq c} \psi_{c_1v\va{k}}^{(m)} \va{x}_{cc_1\va{k}} - \sum_{v_1 \neq v} \psi_{cv_1\va{k}}^{(m)} \va{x}_{v_1v\va{k}} \bigg] \, , \\
		\label{eq:IntrabandPosition}
		&\va{Q}_{nm} \equiv \dfrac{\hbar}{m} \sum_{cv\va{k}} \psi_{cv\va{k}}^{(n)*} \bigg[ \psi_{cv\va{k}}^{(m)} \va{x}_{cc\va{k}} - \psi_{cv\va{k}}^{(m)} \va{x}_{vv\va{k}} \bigg] = i \sum_{cv\va{k}} \psi_{cv\va{k}}^{(n)*} \big[\psi_{cv\va{k}}^{(m)} \big]_{;\va{k}} \, . 
	\end{align}
\end{subequations}
In the last line, the rule $\big( \va{r}_{cc\va{k}}-\va{r}_{vv\va{k}} \big) \psi_{cv\va{k}}^{(m)} = i \big[ \psi_{cv\va{k}}^{(m)} \big]_{;\va{k}}$ has been used \cite{Taghizadeh2017}.

Time reversal symmetry in periodic systems is extremely useful and allows one to choose the phase such that $\psi_{cv(-\va{k})}^{(n)}=\psi_{cv\va{k}}^{(n)*}$, $\va{x}_{nm(-\va{k})}=\va{x}_{mn\va{k}}^*$, and $\va{p}_{nm(-\va{k})}=-\va{p}_{mn\va{k}}^*$. With this choice of phase, one can show that $\va{P}_n=-\va{P}_n^*$, $\va{\Pi}_n=-\va{\Pi}_n^*$, $\va{X}_n=\va{X}_n^*$, $\va{A}_n=-\va{A}_n^*$, $\va{Y}_{nm}=\va{Y}_{nm}^*$, $\va{Q}_{nm}=\va{Q}_{nm}^*$, $\va{P}_{nm}=-\va{P}_{nm}^*$, and $\va{\Pi}_{nm}=-\va{\Pi}_{nm}^*$ \cite{Pedersen2015}. These relations can be used to simplify the expressions of conductivity tensors, which are generally valid for any other phase choice, since all expressions should be independent of the chosen phase.






\end{widetext}

\bibliography{Exciton}

\end{document}